\documentclass[12pt,draftclsnofoot,peerreviewca,onecolumn,letterpaper]{IEEEtran}
\pdfoutput=1

\usepackage{amsmath,amssymb,eucal,graphicx}
\usepackage{epsfig}
\usepackage{exscale}
\usepackage{times}

\long\def\comment#1{}

\newfont{\bbb}{msbm10 scaled 700}

\newfont{\bb}{msbm10 scaled 1100}

\newcommand{\PP}{\mbox{\bb P}}

\newcommand{\EE}{\mbox{\bb E}}

\renewcommand{\arg}{{\hbox{arg}}}

\newcounter{saveeqn}
\newcommand{\alpheqn}{\setcounter{saveeqn}{\value{equation}}%
\stepcounter{saveeqn}\setcounter{equation}{0}%
\renewcommand{\theequation}{\mbox{\arabic{saveeqn}\alph{equation}}}}
\newcommand{\reseteqn}{\setcounter{equation}{\value{saveeqn}}%
\renewcommand{\theequation}{\arabic{equation}}}

\def\BibTeX{{\rm B\kern-.05em{\sc i\kern-.025em b}\kern-.08em
    T\kern-.1667em\lower.7ex\hbox{E}\kern-.125emX}}

\begin{document}

\newtheorem{theorem}{Theorem}
\newtheorem{lemma}{Lemma}
\newtheorem{corollary}{Corollary}
\newtheorem{proposition}{Proposition}
\newtheorem{definition}{Definition}

\title{Hard Fairness Versus Proportional Fairness in
Wireless Communications: \\ The Multiple-Cell Case}
\author{Daeyoung Park, {\it Member, IEEE} and Giuseppe Caire, {\it Fellow, IEEE}
\thanks{D. Park is with the School of Information and Communication Engineering,
Inha University, Incheon, 402-751 Korea. E-mail: dpark@ieee.org.}
\thanks{G. Caire is with the Ming Hsieh
Department of Electrical Engineering, University of Southern
California, Los Angeles, CA 90089 USA. E-mail:
caire@usc.edu.}\thanks{This work was supported by the Korea
Research Foundation Grant funded by the Korean Government (MOEHRD)
(KRF-2006-352-D00141).}} \thispagestyle{empty}

\maketitle

\begin{abstract}
We consider the uplink of a cellular communication system with $K$
users per cell and infinite base stations equally spaced on a
line. The system is conventional, i.e., it does not make use of joint
cell-site processing. A hard fairness (HF) system serves all users with the same rate in any channel
state. In contrast, a system based on proportional fairness serves the users with variable instantaneous rates
depending on their channel state. We compare these two options
in terms of the system spectral efficiency \textsf{C} (bit/s/Hz) versus $E_b/N_0$.
Proportional fair scheduling (PFS) performs generally better than the more restrictive
HF system in the regime of low to moderate SNR, but for high SNR an optimized
HF system achieves throughput comparable to that of
PFS system for finite $K$. The hard-fairness system is interference
limited. We characterize this limit and validate a commonly used simplified model that treats
outer cell interference power as proportional to the in-cell total power and we analytically characterize
the proportionality constant.
In contrast, the spectral efficiency of PFS can grow unbounded for $K \rightarrow \infty$ thanks to
the multiuser diversity effect. We also show that partial frequency/time reuse can
mitigate the throughput penalty of the HF system, especially at high SNR.
\end{abstract}

\newpage
\setcounter{page}{1}

\section{Introduction}

\begin{figure}
\begin{center}
\includegraphics[width=10cm]{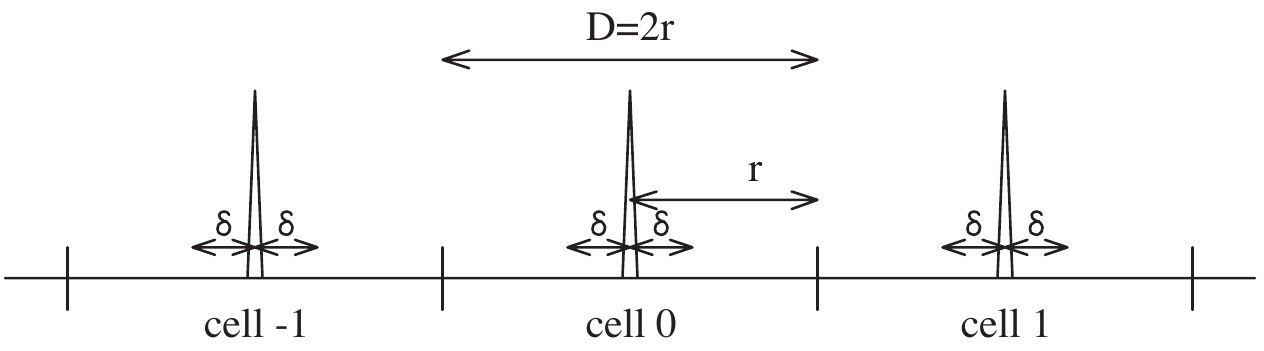}
\end{center}
\caption{Infinite linear array cellular model.}
\label{Fig_infinite_linear_cellular}
\end{figure}

Consider a wireless cellular system with $K$ user terminals (UTs)
per cell where all users share the same bandwidth and
Base Stations (BSs) are arranged on a uniform grid on a line (see Fig.~\ref{Fig_infinite_linear_cellular}).
This model was pioneered by Wyner in
\cite{Wyner94} under a very simplified channel gain assumption,
where the path gain to the closest BS is 1, the
path gain to adjacent BSs is $\alpha$ and it is zero elsewhere.
Wyner considered optimal joint processing of all base stations.
Later, Shamai and Wyner \cite{Shamai97} considered a similar model
with frequency flat fading and more conventional decoding schemes,
ranging from the standard separated base station processing to some forms
of {\em limited cooperation}. A very large literature followed and extended these
works in various ways (see for example \cite{Somekh97,Sanderovich07}).

In this paper we focus on the uplink of a {\em conventional system},
such that each BS decodes only the users in its own cell and treats inter-cell interference
as noise. We extend the model in two directions: 1) we consider realistic propagation channels
determined by a position-dependent path loss,
and by a slowly time-varying frequency-selective fading channel;
2) we compare the {\em optimal} isolated cell delay-limited scheme \cite{Hanly98}
with the Proportional Fair Scheduling (PFS) scheme \cite{Bender00,Tse02}.

In a delay-limited scheme each user transmits at a fixed rate in
each fading block, and the system uses power control in order to
cope with the time-varying channel conditions \cite{Hanly98}. A
delay-limited system achieves ``hard fairness'' (HF), in the sense
that each user transmits at its own desired rate, independently of
the fading channel realization. On the other hand, generally
higher throughput can be achieved by relaxing the fixed rate per slot constraint.
Under variable rate allocation, the sum throughput is maximized by
letting only the user with the best channel transmit in each slot \cite{Tse98}.
However, if users are affected by different distance-dependent path losses
that change on a time-scale much slower than the small scale fading,
this strategy may result in a very unfair resource allocation.
In this case, PFS achieves a desirable tradeoff, by maximizing the geometric
mean of the long-term average throughputs of the users \cite{Tse02}.

HF and PFS have been compared in terms of system throughput versus $E_b/N_0$ in
\cite{Caire07} for the single-cell case.  This comparison is indeed relevant:
HF models how voice-based systems work today
and how Quality-of-Service guaranteed systems  will work in the future
(each user makes a rate request and the system struggles to accommodate it).
In contrast, PFS is being implemented in the so-called EV-DO 3rd
generation systems \cite{Bender00} in order to take advantage of delay-tolerant
data traffic.  Hence, a meaningful question is: {\em what system capacity loss is to be expected by imposing hard-fairness?}
In this paper we address this question by extending the results of \cite{Caire07}
to the multicell case with conventional decoding (i.e., without joint processing of the BSs).

\section{System Model}

Each cell experiences interference from the signals transmitted by UTs
in other cells.  Frequency-selectivity is modeled by considering $M$ parallel
frequency-flat subchannels.  Roughly speaking, we may identify
$M$ with the number of fading coherence bandwidths in the system
signal bandwidth \cite{Tse05,Biglieri98,Proakis01}.
The received signal at BS $n$ in subchannel $m$
is given by
\begin{equation}
r^m (n) = \sum_{j=-\infty}^{\infty} \sum_{k=1}^K h_k^m(n,j)
x_k^m(j) + z^m(n)
\end{equation}
where $h_k^m(n,j)$ denotes the $m$-th subchannel gain from user
$k$ at cell $j$ to cell $n$, and $x_{k}^m(j)$ denotes the signal
of $m$-th subchannel transmitted by user $k$ in cell $j$, and
$z^m(n)$ is an additive white Gaussian noise with variance $N_0$.
The channel (power) gain is given by $g_k^m(n,j)=|h_k^m(n,j)|^2$
and the transmit power of a user is given by
$\EE[|x_k^m(j)|^2]=E_k^m(j)$.

\begin{figure}
\begin{center}
\includegraphics[width=15cm]{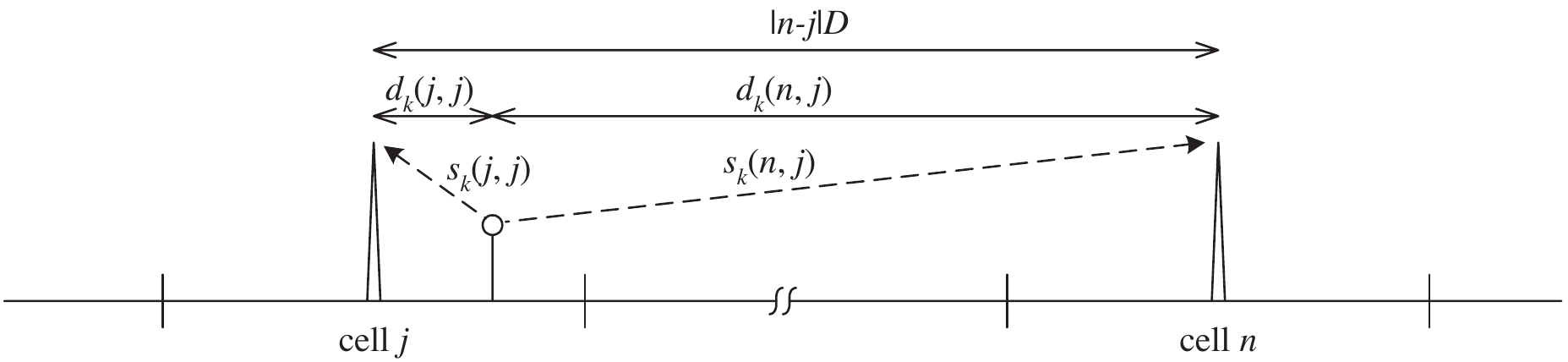}
\begin{center}\small (a) \end{center} 
\includegraphics[width=15cm]{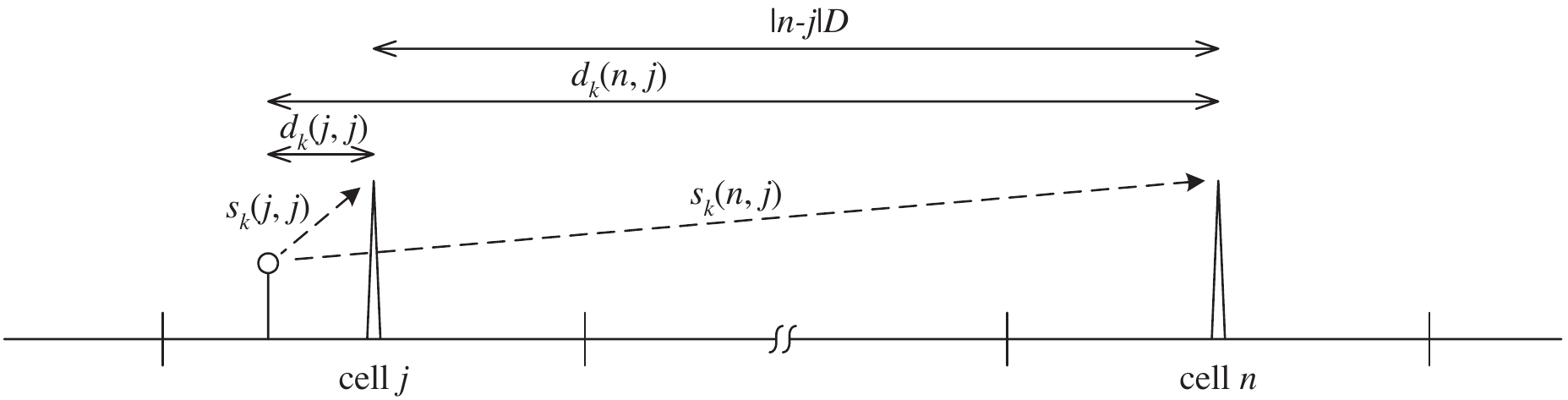}
\begin{center}\small (b) \end{center}
\end{center}
\caption{The path loss $s_k(n,j)$ that is determined by $s_k(j,j)$
in two ways.} \label{Fig_pathloss_model}
\end{figure}

We model the channel gain as the product of two terms, $g_k^m(n,j)=s_k(n,j) f_k^m(n,j)$,
where $s_k(n,j)$ denotes a frequency-flat path gain that depends on
the distance between the BS and the UT, and $f_k^m(n,j)$ is a
``small-scale'' fading term that depends on local scattering environment around
user terminal \cite{Proakis01}.
These two components are mutually independent as they are due to
different propagation effects.
Path loss takes the expression $s_k(n,j) = d_k(n,j)^{-\alpha}$,
where $d_k(n,j)$ denotes the the distance from base station $n$ to
user $k$ in cell $j$ and $\alpha$ is the path loss exponent. We
assume that UTs are not located in a forbidden region at distance less than
$\delta$ from the BS so that the path loss does not diverge.
When the users are uniformly distributed in each cell, the cdf of
$s\equiv s_k(n,n)$ is given by
\begin{equation}
F_s(x) = \left\{ \begin{array}{ll}
  0, & x<r^{-\alpha} \\
  1-\frac{x^{-\frac{1}{\alpha}}-\delta}{r-\delta}, & r^{-\alpha} \leq x < \delta^{-\alpha} \\
  1, & x\geq \delta^{-\alpha}
\end{array} \right.  \label{pathloss_cdf}
\end{equation}
where the cell radius is $r$ and the minimum distance between BSs is $D = 2r$.
The distance $d_k(n,j)$, $n\neq j$, is given by $d_k(n,j) = |n-j|D - d_k(j,j)$
or $d_k(n,j) = |n-j|D + d_k(j,j),$ depending on the location of
user $k$ in cell $j$ as shown in Fig.~\ref{Fig_pathloss_model}.
Consequently, the path loss $s_k(n,j)$ can be expressed as
\begin{equation} s_k(n,j)=\theta_k(n,j)
\left(|n-j|D-s_k(j,j)^{-\frac{1}{\alpha}}\right)^{-\alpha} +
(1-\theta_k(n,j))
\left(|n-j|D+s_k(j,j)^{-\frac{1}{\alpha}}\right)^{-\alpha},
\label{s_k(n,j)}
\end{equation}
where $\theta_k(n,j)$ takes on values 0 or 1 with equal probability since
the UTs are distributed uniformly in each cell.
We assume that the path losses change in time on a very slow scale,
and can be considered as random, but constant, over the whole duration
of transmission. In contrast, the small-scale fading changes relatively rapidly, even
for moderately mobile users \cite{Tse05}. We assume Rayleigh block-fading, changing in an
ergodic stationary manner from block to block, i.i.d., on the $M$ subchannels,
$F_{f}(x)=1-e^{-x}$.

As in \cite{Shamai97}, we assume that all users send independently generated Gaussian random codes.
Let $R_k^m(n)$ denote the rate per symbol allocated by user $k$ in cell $n$ on
subchannel $m$. When outer-cell interference is akin Gaussian noise,
the uplink {\em capacity} region of cell $n$ for fixed channel gains is given by the set of inequalities
\begin{equation}
\sum_{k\in S} R_k^m (n) \leq \log\left(1+\frac{\sum_{k\in S}
g_k^m(n,n) E_k^m(n)}{N_0+I^m(n)} \right) \label{mac_capacity}
\end{equation}
for all $S \subseteq \{1,2,\cdots,K \}$, where the interference at cell $n$ on subchannel $m$ is given by
\begin{equation}
I^m(n) = \sum_{j\neq n}\sum_{k=1}^K g_k^m(n,j) E_k^m(j).
\end{equation}
The capacity region of the $M$ parallel channel case can be
achieved by splitting each user information into $M$ parallel
streams and sending the independent codewords over parallel
channels. The aggregate rate of user $k$ at cell $n$ is given by
\begin{eqnarray}
R_k(n) = \sum_{m=1}^M R_k^m(n),~~~k=1,2,\cdots,K. \label{R_k}
\end{eqnarray}

\subsection{Delay-Limited Systems}

In a delay-limited system, the rates $R_k(n)$ are fixed \emph{a
priori}, and the system allocates the transmit energies in order
such that the rate $K$-tuple $(R_1(n),\cdots,R_K(n))$ is
inside the achievable region in each fading block \cite{Hanly98, Caire07}.
We define the system $E_b/N_0$ under a coding strategy that
supports user rates $(R_1(n),\cdots,R_K(n))$ as
\begin{eqnarray}
\left( \frac{E_b}{N_0} \right)_{\rm sys} = \frac{1}{N_0 \Gamma}
\sum_{k=1}^K \sum_{m=1}^M E_k^m(n), \label{Eb_N0_def}
\end{eqnarray}
where the total number of bits per cell, $\Gamma$, is given by
$\Gamma=\sum_{k=1}^K R_k(n)$. \footnote{Note that we can omit the
cell index $n$ in defining $(E_b/N_0)_{\rm sys}$ as each cell is
symmetric under the assumptions of the infinite linear cellular
model and the symmetric channel distributions.} The system
spectral efficiency $\textsf{C}$ is given by $\textsf{C} =
\frac{\Gamma}{M}$ and it is expressed in bits per second per hertz
(bit/s/Hz) or, equivalently, in bits per dimension.

For given user rates $(R_1(n),\cdots,R_K(n))$, we allocate the
partial rates $R_k^m(n)$ under the constraints (\ref{R_k}) in
order to minimize $(E_b/N_0)_{\rm sys}$. As a subproblem for this
optimization problem, we first consider the $m$-th subchannel
energy allocation assuming that the partial rates and the
interference $I^m(n)$ are given. Thanks to the fact that the
received energy region is a contra-polymatrid \cite{Hanly98}, the
optimal energy allocation is given explicitly as
\begin{eqnarray}
E_{\pi_k^m(n)}^m(n) = \frac{N_0+I^m(n)}{g_{\pi_k^m(n)}^m(n,n)}
\left[\exp\left(\sum_{i\leq k} R_{\pi_i^m(n)}^m(n) \right) -
\exp\left( \sum_{i<k} R_{\pi_i^m(n)}^m(n) \right) \right],
\label{opt_energy}
\end{eqnarray}
where $\pi^m(n)$ is the permutation of $\{1,2,\cdots,K\}$ that
sorts the channel gains in increasing order, i.e.,
$g_{\pi_1^m(n)}^m(n,n) \leq \cdots \leq g_{\pi_K^m(n)}^m(n,n)$ and
the decoding order at base station $n$ is given by $\pi_K^m(n)$
(decoded first), $\pi_{K-1}^m(n), \cdots, \pi_1^m(n)$ (decoded
last).

Since the energy allocation in (\ref{opt_energy}) is the minimum
sum energy allocation for given partial rates, it remains to
minimize the total energy by optimizing the partial rates under the constraints
(\ref{R_k}). Inserting (\ref{opt_energy}) into (\ref{Eb_N0_def}),
we obtain the optimization problem for minimum system $E_b/N_0$
\begin{eqnarray}
\min \frac{1}{N_0 \Gamma} \sum_{k=1}^K \sum_{m=1}^M
\frac{N_0+I^m(n)}{g_{\pi_k^m(n)}^m(n,n)}
\left[\exp\left(\sum_{i\leq k} R_{\pi_i^m(n)}^m(n) \right) -
\exp\left( \sum_{i<k} R_{\pi_i^m(n)}^m(n) \right) \right]
\label{opt_prob}
\end{eqnarray}
under the constraints (\ref{R_k}). The interference experienced by
base station $n$ on subchannel $m$ takes on the expression
\begin{equation}
I^m(n) = \sum_{j\neq n}\sum_{k=1}^K g_{\pi_k^m(n)}^m(n,j)
\frac{N_0+I^m(j)}{g_{\pi_k^m(j)}^m(j,j)}
\left[\exp\left(\sum_{i\leq k} R_{\pi_i^m(j)}^m(j) \right) -
\exp\left( \sum_{i<k} R_{\pi_i^m(j)}^m(j) \right) \right]
\label{interference}
\end{equation}
because the partial rate allocation is performed locally at each
cell. An operating point $(\left  (\frac{E_b}{N_0} \right )_{\rm sys}, \textsf{C})$ on
the power/spectral efficiency plane is a function of both the
signaling strategy and the individual user rates as well as of
the channel gain joint distribution.

\subsection{Delay-Tolerant Systems}

In a delay-tolerant system, the user rates can be adapted
according to their instantaneous channel condition to achieve
higher throughput at the cost of increasing delay. We consider the
constant power allocation and let $\rho=E_{tot}/N_0$ denote the
transmit SNR in each slot. Note that the water-filling power
allocation tends to the constant power allocation as SNR
increases. We also assume that the channel gains are independent
but not necessarily identically distributed across the users, and
symmetrically distributed across the subchannels, that is, for any
permutation $\pi$ of $\{1,2,\cdots,M\}$, the joint cumulative
distribution function of channel gains satisfy
$F(g_k^1(n,j),\cdots,g_k^M(n,j))=F(g_k^{\pi_1}(n,j),\cdots,g_k^{\pi_M}(n,j))$
for all $k$, $n$, $j$. This means that no subchannel is
statistically worse or better than any other.

PFS allocates slots fairly among users in the case of a near-far situation
\cite{Tse02}. The PFS algorithm serves user $k$ on subchannel
$m$ in cell $n$ if $\hat{k}_m(n) = k$, where
\begin{eqnarray}
\hat{k}_m(n) = \arg \max_{k'=1,\cdots,K} \frac{1}{T_{k'}(n)} \log
\left(1+ \frac{\rho g_{k'}^m(n,n)}{1+\rho \sum_{j\neq n}
g_{\hat{k}_m(j)}^m(n,j)} \right). \label{PFS}
\end{eqnarray}
$\hat{k}_m(j)$ denotes the index of the user selected for
transmission on $m$-th subchanel in cell $j$ by the PFS scheduling
and $T_k(n)$ denotes the long-term average user throughput of user
$k$ in cell $n$.

\section{Delay-Limited Systems for a Large Number of Users}

The optimization problem (\ref{opt_prob}) for fixed inter-cell
interference is convex, but it does not yield a closed form
solution. In order to gain insight into the problem we investigate
the asymptotic case for $K \rightarrow \infty$, for which a closed
form solution exists. We make the following assumptions:

\newcounter{steps}
\begin{list}{[A\arabic{steps}]}
 { \usecounter{steps}
  \setlength{\leftmargin}{8.6mm}
  \setlength{\labelwidth}{6.0mm}
  \setlength{\labelsep}{2.7mm}
 }
 \item $M$ is fixed while $K$ becomes arbitrarily large.
 \item As $K\rightarrow \infty$, the empirical distributions of
 $s_k(n,n)$, $\{ f_k^m(n,j) : m = 1,\ldots,M\}$ and of $\theta_k(n,j)$
converge almost surely to given deterministic cdfs,
$F_s(x)$, $F_f(y_1,\cdots,y_M)$, and $F_{\theta}(\phi)$,
respectively.
The cdf $F_f(y_1,\cdots,y_M)$ is symmetric, with identical marginal cdfs $F_f(y)$.
The cdf $F_\theta(\phi)$ has two mass points at 0 and $1$, with equal probability mass $1/2$.

 \item For a given system throughput $\Gamma$, the user individual
rates are given by $R_k(n)=\frac{\Gamma}{K}\nu_k(n)$, where
$\nu_k(n)$ is the rate allocation factor for user $k$ in cell $n$.
As $K\rightarrow \infty$, the empirical rate distribution
converges almost surely to a given deterministic cdf
$F_{\nu}(\mu)$ with mean 1 and support in $[a,b]$, where $0\leq
a\leq b<\infty$ are constants independent of $K$.

 \item The rate allocation factors are fixed \emph{a priori},
independently of the realization of the channel gains. Therefore,
the empirical joint distribution of
$\{(s_k(n,n),f_k^1(n,j),\cdots,f_k^M(n,j),$
$\theta_k(n,j),\nu_k(n) ):k=1,\cdots,K\}$ converges
to the product cdf $F_s(x)F_f(y_1,\cdots,y_M)F_{\theta}(\phi)F_{\nu}(\mu)$.\footnote{
We remark here that this assumption reflects the delay-limited nature
of the problem: the user rates are fixed \emph{a priori} and
independently of the channel realization.}
\end{list}

\subsection{Asymptotic Performance}

In the single cell case, the minimum $(E_b/N_0)_{\rm sys}$ for
given $\textsf{C}$ is given by \cite{Caire07}
\begin{equation}
\left( \frac{E_b}{N_0} \right)_{\rm sys}^{\rm SC} = \log(2) \int
2^{\textsf{C} G_M(x)} \frac{dG_M(x)}{x}.
\label{Eb_N0_SC}
\end{equation}
where $G_M(x)$ denotes the cdf of $s \max\{f^1,\ldots,f^M\}$,
where $s \sim F_s(x)$ and $(f^1,\ldots,f^M) \sim
F_f(y_1,\ldots,y_M)$.  For the infinite linear array cellular
model, we establish the following result.

\vspace{5pt}
\begin{theorem} \label{Theorem1}
Under the assumptions A1, A2, A3, and A4, as $K\rightarrow \infty$
the minimum $(E_b/N_0)_{\rm sys}$ for given system spectral
efficiency $\textsf{C}$ in an infinite linear array cellular model
is given by
\begin{equation}
\left( \frac{E_b}{N_0} \right)_{\rm sys}^{\rm MC} = \frac{\log(2)
\int 2^{\textsf{C} G_M(x)}
\frac{d G_M(x)}{x}} {1-\textsf{C} \log(2) \iint
2^{\textsf{C} G_M(xy)} \Phi(x) \frac{dF_s(x)
dH_M(y)}{xy} }. \label{Eb_N0_MC}
\end{equation}
where $H_M(y)$ denotes the cdf of $\max\{f^1,\ldots,f^M\}$ where
$(f^1,\ldots,f^M) \sim F_f(y_1,\ldots,y_M)$, and where the
function $\Phi(s)$ is given by
\begin{equation}
\Phi(x) =
D^{-\alpha}\left(\zeta\left(\alpha,1-\frac{x^{-1/\alpha}}{D}\right)
+\zeta\left(\alpha,1+\frac{x^{-1/\alpha}}{D}\right) \right),
\label{Psi_func}
\end{equation}
where $\zeta(a,x)$ is the Riemann zeta function
\cite{Gradshteyn00}.
The minimum $(E_b/N_0)_{\rm sys}$ is achieved
by letting each user transmit on its best subchannel only, and by
using superposition coding and successive decoding on each
subchannel.
\end{theorem}
\begin{proof}
See Appendix A.
\end{proof}
\vspace{5pt}

If we compare (\ref{Eb_N0_MC}) with (\ref{Eb_N0_SC}), we can
observe three facts: First, $(E_b/N_0)_{\rm sys}^{\rm MC}$ is
higher than $(E_b/N_0)_{\rm sys}^{\rm SC}$ for all $\textsf{C}$.
So, for the multiple cell case, the higher $(E_b/N_0)_{\rm sys}$
is required to achieve the same system spectral efficiency
$\textsf{C}$ due to intercell interference. Second,
$(E_b/N_0)_{\rm sys}^{\rm MC}$ converges to $(E_b/N_0)_{\rm
sys}^{\rm SC}$, as $\textsf{C}\rightarrow 0$. For low spectral
efficiency, low transmit energy causes negligible interference to
other cells, which effectively turns the multiple cell case into
the single cell (interference-free) case. Third, there exists a
spectral efficiency limit in the multiple cell case because
$(E_b/N_0)_{\rm sys}^{\rm SC}$ tends to infinity as $\textsf{C}
\rightarrow \textsf{C}_0$, where $\textsf{C}_0$ is the root of the
denominator of the right hand side in (\ref{Eb_N0_MC})
\begin{equation}
1-\textsf{C}_0 \log(2) \iint 2^{\textsf{C}_0 G_{M}(xy)}
\Phi(x) \frac{dF_s(x) dH_M(y)}{xy}=0. \label{C_0}
\end{equation}
Any spectral efficiency can be supported in the single cell case
as long as the transmit energy is high. However, in the multiple
cell case, high transmit energy to support high spectral
efficiency may cause significant interference to other cells, thus
demanding more transmit energy to combat interference from other
cells. Eventually, additional energy to combat interference
becomes tremendously large even in a finite spectral efficiency.
So, the multiple cell case is an \emph{interference-limited}
system due to this fundamental spectral efficiency limit.

\begin{figure}
\begin{center}
\includegraphics[width=10cm]{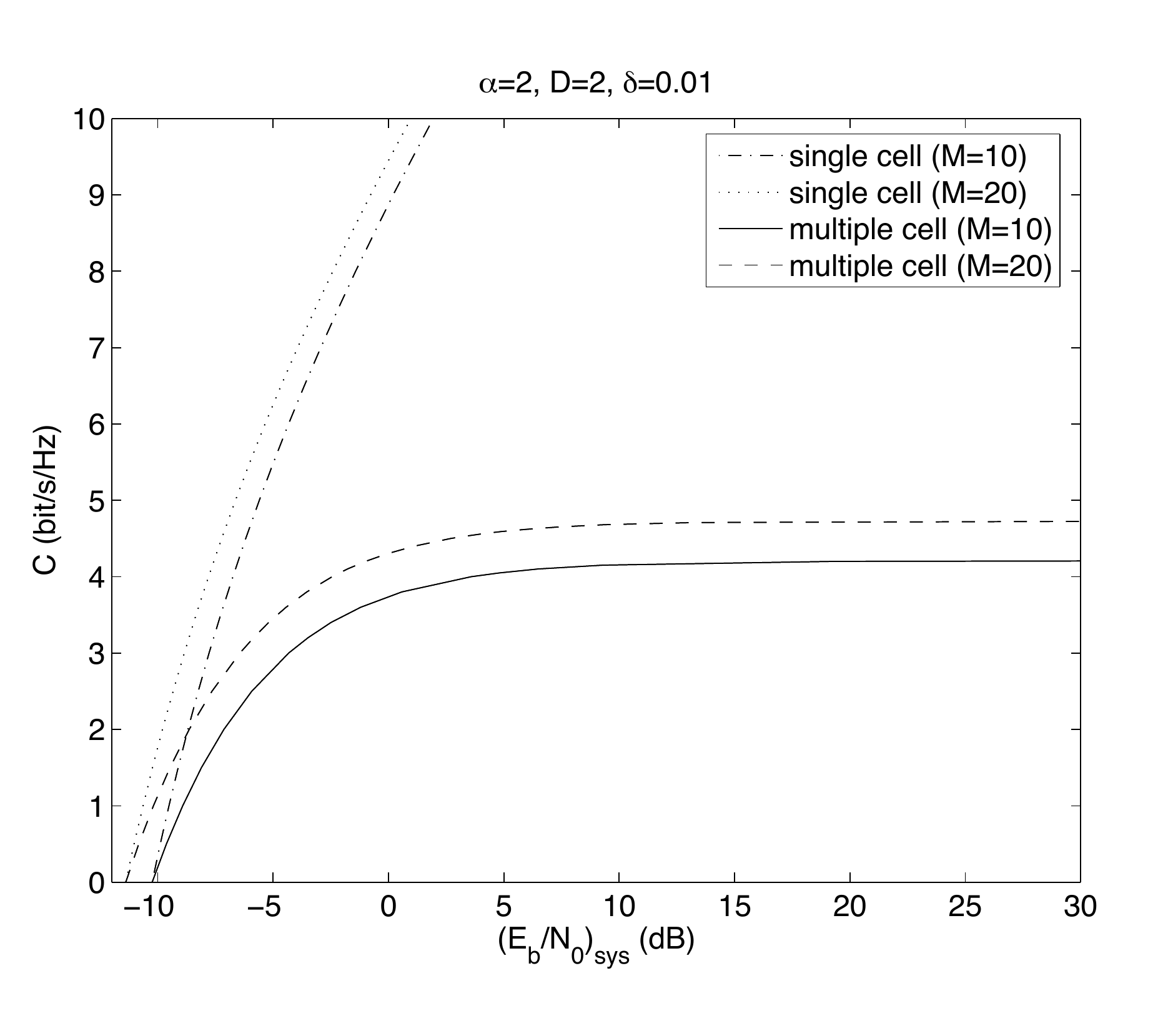}
\end{center}
\caption{Spectral efficiency versus system $E_b/N_0$ for the
optimal delay-limited systems for $K=\infty$. The channel
parameters are $M=10$, the path loss exponent $\alpha=2$, the cell
size $D=2$, and the forbidden region $\delta=0.01$.}
\label{Fig_capacity_HFS}
\end{figure}

Fig.~\ref{Fig_capacity_HFS} shows the spectral efficiency achieved
by the delay-limited systems for infinite number of users and we
can confirm the three properties described above. The
spectral efficiency limits for $M=10$ and $M=20$ are
$\textsf{C}_0=4.2$ bits/s/Hz and $\textsf{C}_0=4.73$ bits/s/Hz,
respectively.

\subsection{Asymptotic Results for Simplified Multiple Cell Model}

A well-known approach to the computation of system
capacity of a conventional multi-cell
system consists of modeling the inter-cell interference power as
proportional to the total transmit power in each cell
\cite{Viterbi95}. In this section we validate this approach and
compute explicitly the proportionality constant. Let the
interference level experienced by each cell be given by
\begin{equation}
I^m = \beta \sum_{k=1}^K E_k^m. \label{simple_interference}
\end{equation}
where we drop the cell index $n$ for simplicity. In
(\ref{simple_interference}), the ratio of interference and total
transmit power is given by $\beta$. The capacity region of cell
$n$ for cell-site optimal joint decoding is given by
\begin{equation}
\sum_{k\in S} R_k^m \leq \log\left(1+\frac{\sum_{k\in S} g_k^m
E_k^m}{N_0+I^m} \right) \label{simple_mac_capacity}
\end{equation}
for all $S \subseteq \{1,2,\cdots,K \}$. For given $I^m$, the
minimum total energy supporting a given rate ${\bf
R}^m=(R_1^m,\cdots,R_K^m)$ with gains ${\bf
g}^m=(g_1^m,\cdots,g_K^m)$ is given by
\begin{eqnarray}
E_{\pi_k^m}^m = \frac{N_0+I^m}{g_{\pi_k^m}^m}
\left[\exp\left(\sum_{i\leq k} R_{\pi_i^m}^m \right) - \exp\left(
\sum_{i< k} R_{\pi_i^m}^m \right) \right],
\label{simple_opt_energy}
\end{eqnarray}
where, as before, $\pi^m$ is sorting permutation of the channel gains in increasing order.
We have:

\vspace{5pt}
\begin{theorem} \label{Theorem2}
Under the assumptions A1, A2, A3, and A4, as $K\rightarrow \infty$
the minimum $(E_b/N_0)_{\rm sys}$ for given system spectral
efficiency $\textsf{C}$ in a simple multiple cell model is given by
\begin{equation}
\left( \frac{E_b}{N_0} \right)_{\rm sys}^{\rm MC} = \frac{\log(2)
\int 2^{\textsf{C} G_{M}(x)}
\frac{dG_M(x)}{x}} {1- \beta \textsf{C} \log(2) \int
2^{\textsf{C} G_M(x)}\frac{dG_M(x)}{x}}
\label{simple_Eb_N0_MC}
\end{equation}
The minimum $(E_b/N_0)_{\rm sys}$ is achieved by letting each user
transmit on its best subchannel only, and by using superposition
coding and successive decoding on each subchannel.
\end{theorem}
\begin{proof}
See Appendix C.
\end{proof}
\vspace{5pt}

\begin{figure}
\begin{center}
\includegraphics[width=10cm]{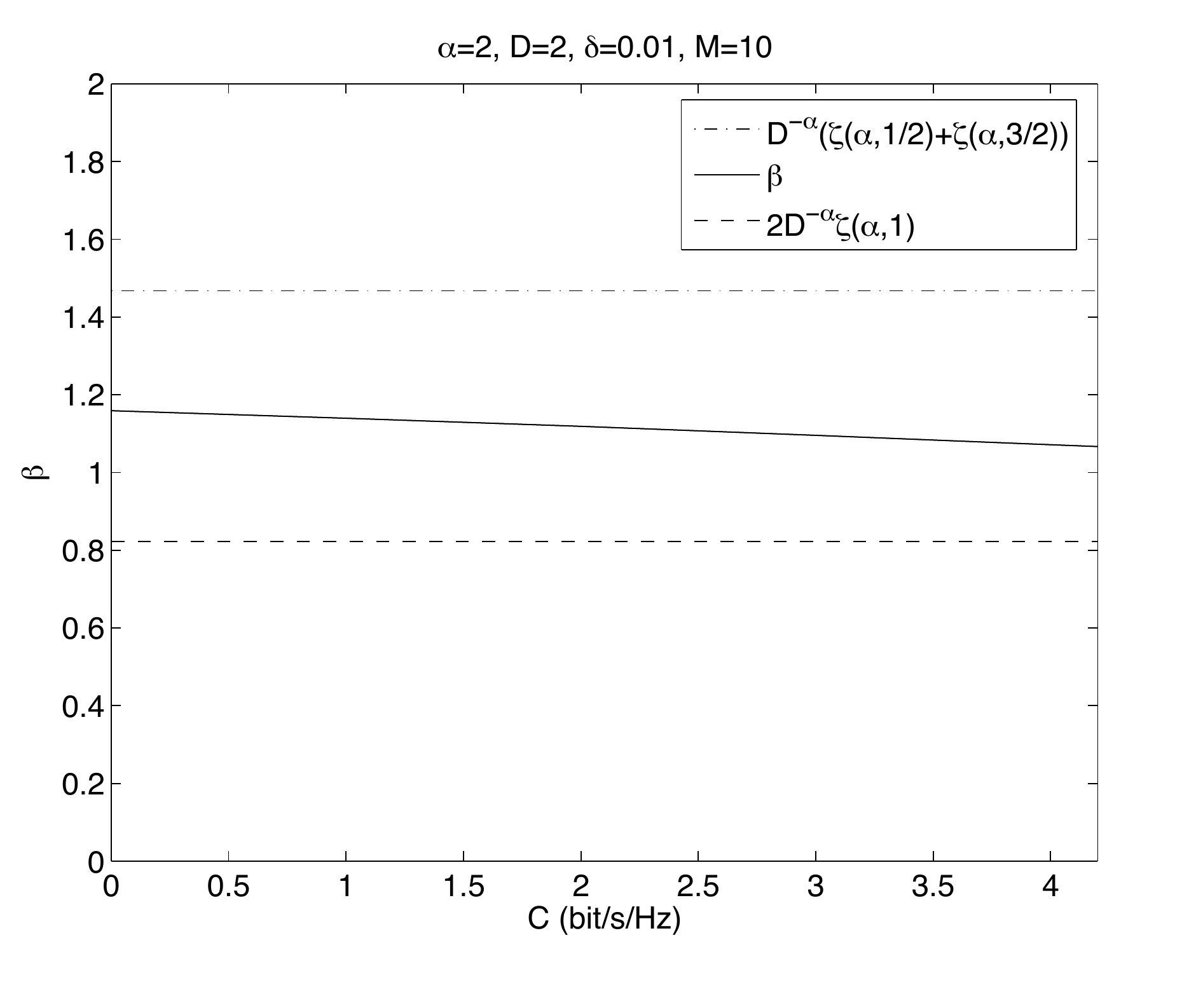}
\end{center}
\caption{Power fraction $\beta$ and its upper and lower bounds for
the optimal delay-limited systems for $K=\infty$. The channel
parameters are path loss exponent $\alpha=2$, the cell size $D=2$,
and the forbidden region $\delta=0.01$.} \label{Fig_beta}
\end{figure}

By equating (\ref{Eb_N0_MC}) and (\ref{simple_Eb_N0_MC}), we can
solve for $\beta$ and obtain
\begin{equation}
\beta = \frac{\iint 2^{\textsf{C} G_M(xy)} \Phi(x)
\frac{dF_s(x) dH_M(y)}{xy}} {\int 2^{\textsf{C}
G_M(x)}\frac{dG_M(x)}{x}}.
\label{eq_beta}
\end{equation}
Since $\zeta(a,1-x)+\zeta(a,1+x)$ is an increasing function of $x$
and $0 \leq s^{-1/\alpha}/D \leq 1/2$, we have
\begin{equation}
2D^{-\alpha} \zeta(\alpha,1)\leq \beta \leq D^{-\alpha}
(\zeta(\alpha,1/2)+\zeta(\alpha,3/2)). \label{beta_bound}
\end{equation}
For example, if $\alpha=2$, then $3.29D^{-2} \leq \beta \leq 5.87
D^{-2}$. This implies that there exists $\beta$ in $[2D^{-\alpha}
\zeta(\alpha,1), D^{-\alpha}
(\zeta(\alpha,1/2)+\zeta(\alpha,3/2))]$ such that the simple
multiple cell model yields a valid result as in the infinite
linear array cellular model. Fig.~\ref{Fig_beta} shows $\beta$
in (\ref{eq_beta}) and its upper and lower bounds in
(\ref{beta_bound}) for $M=10$, path loss exponent $\alpha=2$, the
cell size $D=2$, and the forbidden region $\delta=0.01$. We
observe that $\beta$ changes very little with respect to the spectral
efficiency \textsf{C}. Hence, assuming $\beta$ constant as commonly done
in simplified multicell analysis \cite{Viterbi95} yields good approximations.

We can rewrite the minimum $(E_b/N_0)_{\rm sys}^{\rm MC}$ in terms
of $(E_b/N_0)_{\rm sys}^{\rm SC}$
\begin{equation}
\left( \frac{E_b}{N_0} \right)_{\rm sys}^{\rm MC} = \frac{\left(
\frac{E_b}{N_0} \right)_{\rm sys}^{\rm SC}} {1-\beta \textsf{C}
\left( \frac{E_b}{N_0} \right)_{\rm sys}^{\rm SC}}.
\end{equation}
So, the performance degradation due to the multiple cell
interference is up to 3 dB as long as
\begin{equation}
\left( \frac{E_b}{N_0} \right)_{\rm sys}^{\rm SC} \leq
\frac{1}{2\beta \textsf{C}}. \label{noise_limited_condition}
\end{equation}
In the case when the interference dominates the noise, the
performance degradation is more than 3 dB and the $(E_b/N_0)_{\rm
sys}^{\rm MC}$ tends to infinity as $\textsf{C} \rightarrow
\textsf{C}_0$. We may summarize the conditions that determine the
performance of the multiple cell case
\begin{eqnarray}
\left\{ \begin{array}{cl}
 \left( \frac{E_b}{N_0} \right)_{\rm sys}^{\rm SC} \leq
\frac{1}{2\beta \textsf{C}},&\mbox{noise dominated region} \\
 \frac{1}{2\beta \textsf{C}} < \left( \frac{E_b}{N_0} \right)_{\rm sys}^{\rm SC}
< \frac{1}{\beta \textsf{C}},&\mbox{interference dominated region} \\
 \left( \frac{E_b}{N_0} \right)_{\rm sys}^{\rm SC} \geq \frac{1}{\beta \textsf{C}},&\mbox{forbidden region}
\end{array} \right. \label{condition_for_mc}
\end{eqnarray}
For $\textsf{C}=2.8$ bit/s/Hz, $\left( \frac{E_b}{N_0}
\right)_{\rm sys}^{\rm SC}=-7.967$ dB and $\frac{1}{2\beta
\textsf{C}}=-7.896$ dB when $\beta\simeq 1.1$ (see
Fig.~\ref{Fig_beta}). According to (\ref{condition_for_mc}), the
system may be regarded as operated in the noise-dominated region
because $\left( \frac{E_b}{N_0} \right)_{\rm sys}^{\rm
SC}<\frac{1}{2\beta \textsf{C}}$. We can verify this by checking
$\left( \frac{E_b}{N_0} \right)_{\rm sys}^{\rm MC}=-5.024$ dB in
Fig.~\ref{Fig_capacity_HFS}. So, as long as the target spectral
efficiency is less than 2.8 bit/s/Hz, additionally required energy
in the multiple cell is no more than 3 dB compared to the single
cell.\footnote{We remark here that the values of $E_b/N_0$ should
be considered on a relative scale: a horizontal dB shift of all
these curves can be obtained simply by rescaling the cell size $D$
and by introducing a path-loss normalization factor. However, the
relevant information here is captured by the relative values
(differences in dB) and by the spectral efficiency.} It should
also be noticed that practical systems nowadays achieve spectral
efficiencies of about 1 bit/s/Hz. It follows that, in practice,
there is still a significant gap before the interference-limited
nature of the multicell system becomes significant. Therefore,
implementing some clever low complexity separated cell-site
processing may not be a bad idea from the practical system
engineering viewpoint. For example, using successive decoding at
each BS, as the system analyzed in this paper, represents already
a remarkable step-forward with respect to orthogonal or
semi-orthogonal conventional techniques such as TDMA, FDMA, CDMA,
OFDM or a mixture thereof.

\subsection{Partial Reuse Transmission}

In order to alleviate the interference limited nature of the multi-cell delay-limited system,
we introduce a partial reuse transmission scheme. We notice here that in the traditional Wyner cellular model
non-adjacent do not interfere. Trivially, a reuse factor of 2 (even-odd cells)
yields a non-interference limited system \cite{Shamai97}. With a more realistic distance-dependent path loss model
as in our case, this is no longer true and reuse factor optimization becomes more delicate.

\begin{figure}
\begin{center}
\includegraphics[width=15cm]{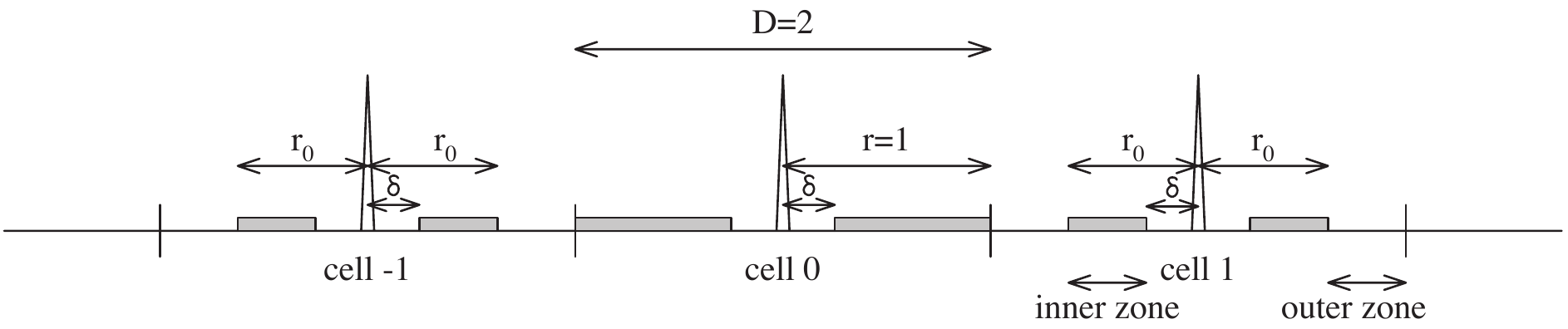}
\begin{center}\small (a) \end{center} 
\includegraphics[width=15cm]{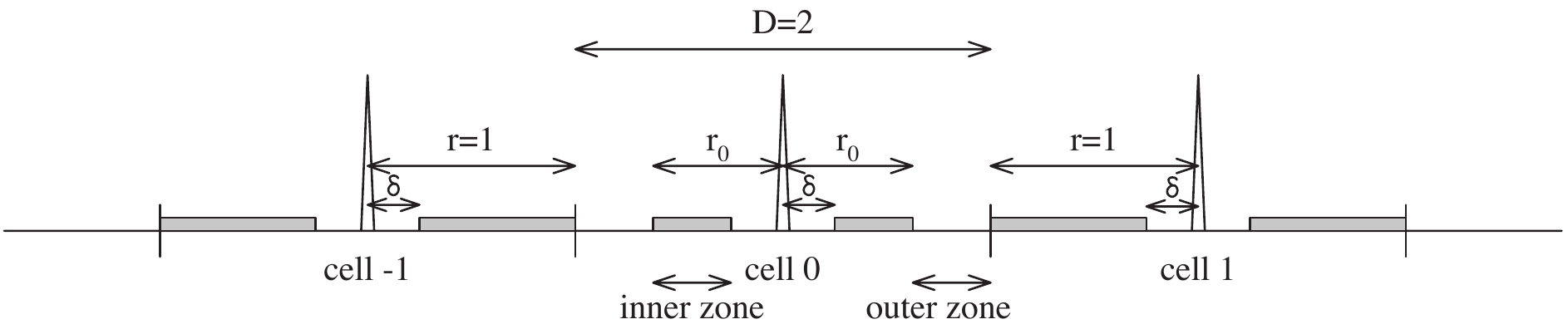}
\begin{center}\small (b) \end{center}
\end{center}
\caption{Cellular model for partial reuse transmission in which
users located in shaded areas are only allowed to transmit signals
in (a) phase 1 and (b) phase 2. } \label{Fig_partial_reuse}
\end{figure}

Without loss of generality we can set the cell size $D=2$ (generalization is trivially obtained by cell re-scaling).
Fig.~\ref{Fig_partial_reuse} shows the system model for
partial reuse transmission. We classify cells into even and odd groups, according to their index parity.
We also divide each cell into an inner and outer regions, where the inner region radius is $\delta \leq r_0 \leq 1$.
The users located in the inner zone transmit in each slot. Users located in the
outer zone transmit signals alternately, at even slot times if they belong to an even cell, or odd slot times if they belong to an odd cell.  Consequently, the activity duty cycle of the users located in the outer zone
is 0.5.

From (\ref{pathloss_cdf}),  the cdf of the path loss when users
are uniformly distributed in $(a, b)$ is given by
\begin{equation}
F_s^{(a,b)}(x) = \left\{ \begin{array}{ll}
  0, & x<b^{-\alpha} \\
  1-\frac{x^{-\frac{1}{\alpha}}-a}{b-a}, & b^{-\alpha} \leq x < a^{-\alpha} \\
  1, & x\geq a^{-\alpha}
\end{array}. \right.  \label{pathloss_cdf_(a,b)}
\end{equation}
Due to the symmetry, it suffices to focus only on the phase where
even cells are fully active. In this case, the active users in odd
cells are uniformly distributed in $(\delta, r_0)$ while the users
in even cells are uniformly distributed in $(\delta, 1)$. Letting
$K_0 = K$ denote the number of active users in even cells, the
number of active users in odd cells is
$K_1=\frac{r_0-\delta}{1-\delta}K$. Similarly, we set the number
of bits transmitted in each odd cell is
$\Gamma_1=\frac{r_0-\delta}{1-\delta}\Gamma_0$ when the number of
bits transmitted in each even cell is $\Gamma_0$. Based on these,
the spectral efficiency in this partial reuse system is
\begin{equation}
\textsf{C} =
\frac{\Gamma_0+\Gamma_1}{2M}=\frac{\Gamma_0}{2M}\frac{1+r_0-2\delta}{1-\delta}.
\label{C_partial}
\end{equation}
Also, the $(E_b/N_0)_{\rm sys}$ is defined by
\begin{equation}
\left(\frac{E_b}{N_0}\right)_{\rm sys} = \frac{E_{\rm
tot}^0+E_{\rm tot}^1}{N_0 (\Gamma_0+\Gamma_1)} = \frac{E_{\rm
tot}^0+E_{\rm tot}^1}{2N_0 M \textsf{C}} ,
\label{Eb_N0_partial_def}
\end{equation}
where $E_{\rm tot}^0$ and $E_{\rm tot}^1$ denote the total
transmitted energy in each even cell and each odd cell,
respectively. We have:

\vspace{5pt}
\begin{theorem}  \label{Theorem_HF_Partial}
Under the assumptions A1, A2, A3, and A4, as $K\rightarrow
\infty$, the minimum $(E_b/N_0)_{\rm sys}$ for given system
spectral efficiency $\textsf{C}$ for the partial reuse system in
an infinite linear array cellular model is given by
\begin{eqnarray}
\left( \frac{E_b}{N_0} \right)_{\rm sys}^{\rm MC-Partial} =
\frac{1}{2}
\frac{1+A_{01}-A_{11}}{1-A_{00}-A_{11}-A_{01}A_{10}+A_{01}A_{11}}
\frac{2(1-\delta)}{1+r_0-2\delta} \nonumber \\ \cdot \log(2) \int
2^{\frac{2\textsf{C}(1-\delta)}{1+r_0-2\delta}
G_M(x)}
\frac{dG_M(x)}{x}~~ \nonumber \\
+\frac{1}{2}
\frac{1+A_{10}-A_{00}}{1-A_{00}-A_{11}-A_{01}A_{10}+A_{01}A_{11}}
\frac{2(r_0-\delta)}{1+r_0-2\delta} \nonumber \\ \cdot \log(2) \int
2^{\frac{2\textsf{C}(r_0-\delta)}{1+r_0-2\delta}
G_M^{(\delta,r_0)}(x)}
\frac{dG_M^{(\delta,r_0)}(x)}{x}~.
\label{E_b/N_0_HF_Partial}
\end{eqnarray}
Here $A_{ij}$'s are given by \alpheqn
\begin{eqnarray}
A_{00} = \log(2) \frac{2\textsf{C}(1-\delta)}{1+r_0-2\delta} \iint
2^{\frac{2\textsf{C}(1-\delta)}{1+r_0-2\delta}
G_M(xy)} \Phi_0(x)
\frac{dF_s(x) dH_M(y)}{xy}~  \label{eq30a} \\
A_{01} = \log(2) \frac{2\textsf{C}(r_0-\delta)}{1+r_0-2\delta} \iint
2^{\frac{2\textsf{C}(r_0-\delta)}{1+r_0-2\delta}
G_M^{(\delta,r_0)}(xy)} \Phi_1(x)
\frac{dF_s^{(\delta,r_0)}(x) dH_M(y)}{xy}  \label{eq30b} \\
A_{10} = \log(2) \frac{2\textsf{C}(1-\delta)}{1+r_0-2\delta} \iint
2^{\frac{2\textsf{C}(1-\delta)}{1+r_0-2\delta}
G_M(xy)} \Phi_1(x)
\frac{dF_s(x) dH_M(y)}{xy}~  \label{eq30c} \\
A_{11} = \log(2) \frac{2\textsf{C}(r_0-\delta)}{1+r_0-2\delta} \iint
2^{\frac{2\textsf{C}(r_0-\delta)}{1+r_0-2\delta}
G_M^{(\delta,r_0)}(xy)} \Phi_0(x)
\frac{dF_s^{(\delta,r_0)}(x) dH_M(y)}{xy} \label{eq30d}
\end{eqnarray} \reseteqn
where \alpheqn
\begin{eqnarray}
\Phi_0(x) =
4^{-\alpha}\left(\zeta\left(\alpha,1-\frac{x^{-1/\alpha}}{4}\right)
+\zeta\left(\alpha,1+\frac{x^{-1/\alpha}}{4}\right) \right)~~
 \\
\Phi_1(x) =
4^{-\alpha}\left(\zeta\left(\alpha,\frac{1}{2}-\frac{x^{-1/\alpha}}{4}\right)
+\zeta\left(\alpha,\frac{1}{2}+\frac{x^{-1/\alpha}}{4}\right)
\right).
\end{eqnarray} \reseteqn
\end{theorem}
\vspace{5pt}
\begin{proof}
See Appendix D.
\end{proof}
\vspace{5pt}

In the expressions above, $G_M^{(\delta,r_0)}(x)$ denotes the cdf
of $s \max\{f^1,\ldots,f^M\}$ when $s \sim F_s^{(\delta,r_0)}(x)$
and $(f^1,\ldots,f^M) \sim F_f(y_1,\ldots,y_M)$. It can be readily
checked that the case of $r_0=1$ coincides with the full reuse,
that we have analyzed separately in Theorem \ref{Theorem1}. If
$r_0=\delta$, then half of cells are silent to reduce
interference, which corresponds to the reuse factor of 2, also
referred to as intercell time division transmission system. We
also notice that the performance of frequency reuse, i.e.,
partitioning the $M$ frequency subchannels into two subsets and
allocate them to even and odd cells, cannot outperform the
``time'' reuse studied in Theorem \ref{Theorem_HF_Partial} since
the frequency diversity in each cell would be reduced. Of course,
the reuse parameter $r_0$ may be optimized in order to minimize
the required $(E_b/N_0)_{\rm sys}$ for given spectral efficiency
\textsf{C}.

\begin{figure}
\begin{center}
\includegraphics[width=10cm]{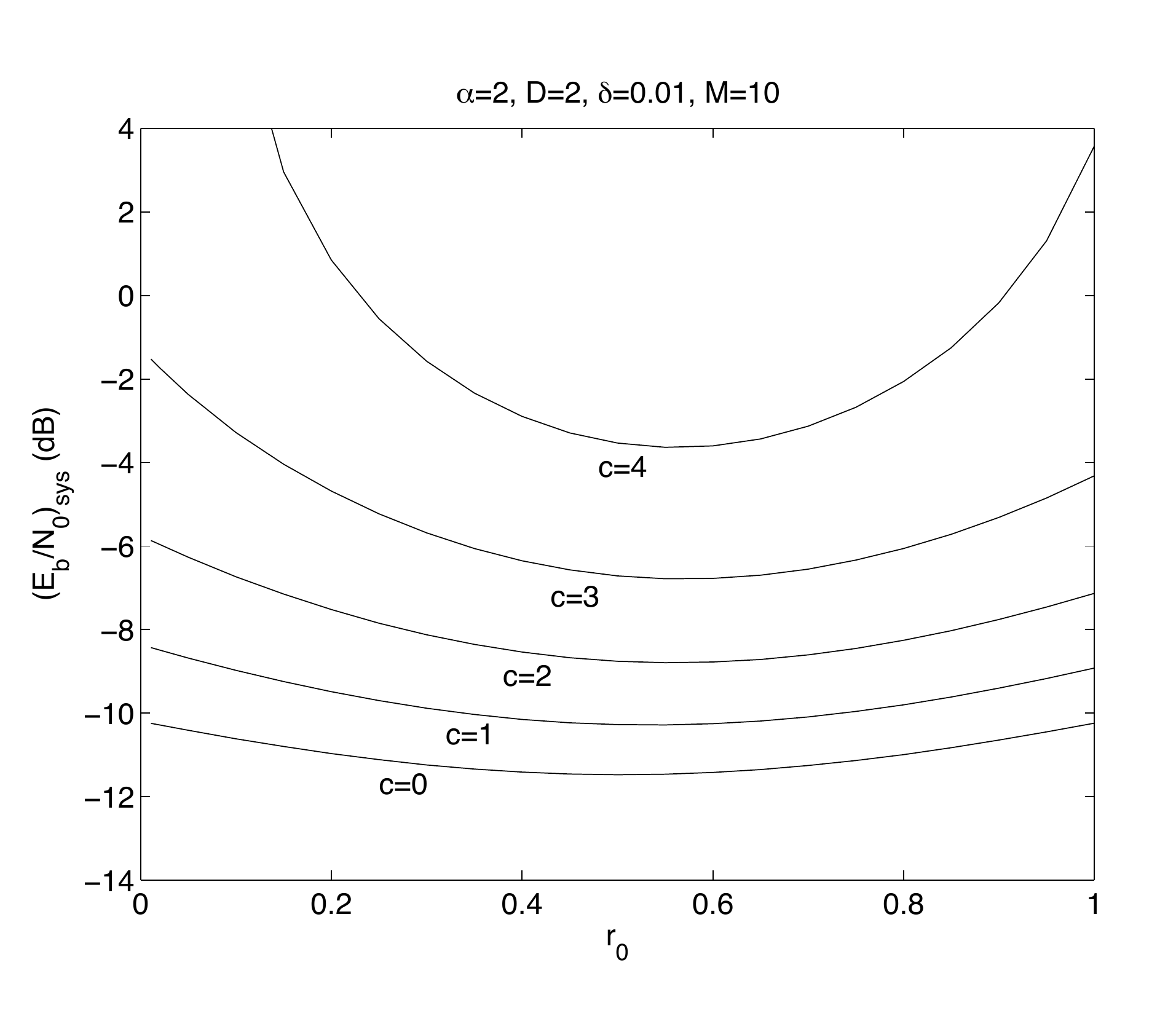}
\end{center}
\caption{System $E_b/N_0$ versus $r_0$ for the delay-limited
system in the partial reuse transmission scheme for $K=\infty$.
The channel parameters are the path loss exponent $\alpha=2$, the
cell size $D=2$, the forbidden region $\delta=0.01$, and $M=10$.}
\label{Fig_EbN0_partial_r_0}
\end{figure}

Fig.~\ref{Fig_EbN0_partial_r_0} depicts $(E_b/N_0)_{\rm sys}^{\rm
MC-Partial}$ in (\ref{E_b/N_0_HF_Partial}) versus $r_0$ for the
partial reuse transmission system. There is an optimal $r_0 \in
(\delta , 1]$ that minimizes $(E_b/N_0)_{\rm sys}^{\rm
MC-Partial}$ for given \textsf{C}. At $\textsf{C}=4$, the
$(E_b/N_0)_{\rm sys}$ difference between the full transmission and
the optimal partial reuse transmission is about 7 dB, that is
quite significant. This shows that large gains can be realized by
careful optimization of the partial reuse factor.

\begin{figure}
\begin{center}
\includegraphics[width=10cm]{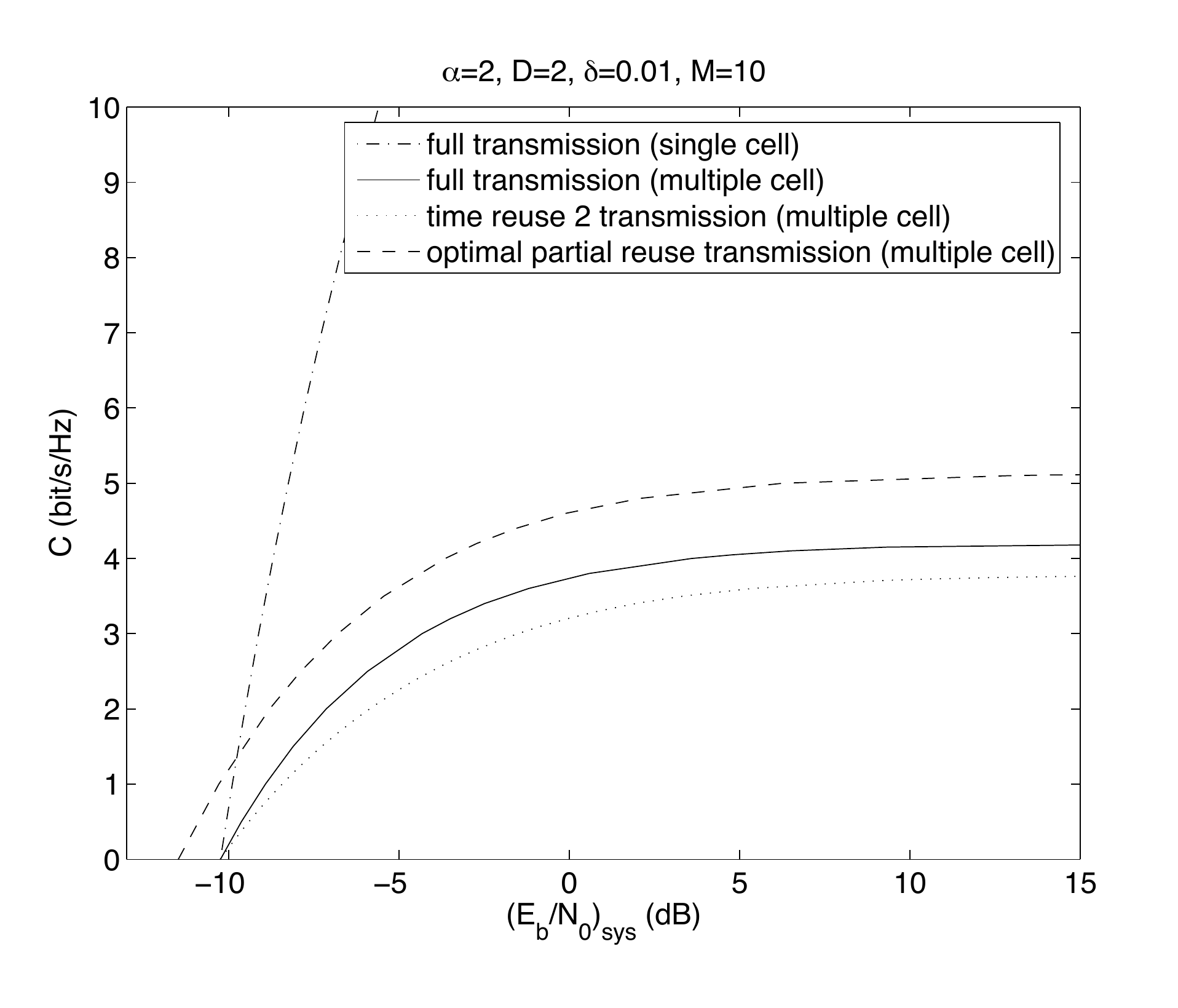}
\end{center}
\caption{Spectral efficiency versus system $E_b/N_0$ for the
delay-limited system in the optimal partial reuse transmission
scheme for $K=\infty$. The channel parameters are the path loss
exponent $\alpha=2$, the cell size $D=2$, the forbidden region
$\delta=0.01$, and $M=10$.} \label{Fig_EbN0_partial_opt}
\end{figure}

Fig.~\ref{Fig_EbN0_partial_opt} shows the optimal partial reuse
transmission scheme compared with the full transmission schemes in
the single and multiple cell cases and reuse 2 scheme
in the multiple cell case.
We notice here that the gain due to partial reuse comes at the expenses of relaxing slightly the
HF constraint: users in the outer zone of each cell is half of their requested rate,
since they are served with duty cycle 0.5. For this reason, in the regime of small $\textsf{C}$ the multicell system with partial reuse
outperforms the single-cell HF system.

\section{Performance of Delay-Tolerant Systems}

A delay-tolerant system can achieve high spectral efficiency at
the cost of loose delay requirement.
In a distance-dependent path-loss scenario as considered here, it can be shown that PFS  serves
user at the peak of its own small-scale fading, i.e., the path-loss takes no role in the channel allocation
\cite{Caire07}. In this section, we investigate the performance of PFS in our multi-cell setting.

\subsection{Spectral Efficiency Bounds}

According to Theorem 1 in \cite{Caire07}, in the case of the
single cell, the spectral efficiency \textsf{C} as a function of
$\left( \frac{E_b}{N_0} \right)_{\rm sys}$ is given implicitly by
\begin{eqnarray}
\textsf{C}=\EE \left[ \log_2(1+\rho s  \max_k f_k )\right] \nonumber \\
\left( \frac{E_b}{N_0} \right)_{\rm sys}^{\rm SC-PFS} =
\frac{\rho}{\textsf{C}}~~~~~~~~~~~~~~~~~~~~~~~~~~~
\end{eqnarray}
where $s \sim F_s(x)$ given by
(\ref{pathloss_cdf}) and $f_k$ is distributed as the frequency
selective block fading of user $k$.
Similarly, we have:

\vspace{5pt}
\begin{theorem}
For any given $K$, the spectral efficiency and the system
$E_b/N_0$ achieved by the PFS are given by
\begin{eqnarray}
\textsf{C}=\EE \left[ \log_2\left(1+ \frac{\rho s  \max_{k=1,2,\cdots,K} f_k}{1+\rho \sum_{j\neq n} g_{\hat{k}(j)}(n,j)} \right)\right] \nonumber \\
\left( \frac{E_b}{N_0} \right)_{\rm sys}^{\rm MC-PFS} =
\frac{\rho}{\textsf{C}}~~~~~~~~~~~~~~~~~~~~~~~~~~~~~~~~~~~~~~~~~~~~~
\label{Eb_N0_PFS}
\end{eqnarray}
where $\hat{k}(j)$ is the index of the user selected for transmission in cell $j$.
\end{theorem}
\vspace{5pt}
\begin{proof}
Because of the symmetry of the small-scale fading distribution,
the average throughput for each subchannel is identical. So, it is
sufficient to focus on a single subchannel and we drop the
subchannel index $m$. Since each user in the same cell gets the
same interference $I=\sum_{j\neq n} g_{\hat{k}(j)}(n,j)$ in
(\ref{PFS}), the PFS scheduling decision in the multiple cell is
the same with that in the single cell if the noise is replaced
with $1+\rho I$. From Theorem 1 in \cite{Caire07}, the result
follows straightforwardly.
\end{proof}
\vspace{5pt}

From (\ref{s_k(n,j)}), $g_{\hat{k}(j)}(n,j)$ can be expressed as
\begin{equation}
g_{\hat{k}(j)}(n,j) \sim \left( \theta (|n-j|D-U)^{-\alpha} +
(1-\theta) (|n-j|D+U)^{-\alpha} \right) f(n,j),
\label{interf_dist}
\end{equation}
where $U$ is a uniform random variable ranging from $\delta$ to
$r$, $\theta$ is a binary random variable taking 0 or 1 with equal
probability, and $f(n,j)$ is unit-mean exponentially distributed.

We can derive some bounds for the spectral efficiency as shown in
Appendix E. The lower bound follows from Jensen's inequality and is given by
\begin{eqnarray}
\textsf{C} \geq \EE \left[ \log_2\left(1+ \frac{\rho s  \max_k
f_k}{1+\rho I_0} \right)\right], \label{PFS_lower_bound}
\end{eqnarray}
where the average interference is given by
\begin{eqnarray}
I_0=\EE\left[ \sum_{j\neq n} g_{\hat{k}(j)}(n,j) \right]~~~~~~~~~~~~~~~~~~~~~~~~~~~~~~~~~~~~~~~~~~~~~~~~~~~~~~~~~~~~~~~~~ \nonumber \\
= \left\{
\begin{array}{ll}
 \frac{1}{D(D/2-\delta)} \left(2-\frac{D}{\delta}+\pi\cot\left(\frac{\pi\delta}{D} \right) \right), & \mbox{if }\alpha=2 \\
 \frac{D^{-\alpha+1}}{(\alpha-1)(D/2-\delta)}\left(2^{\alpha-1}
+\zeta\left(\alpha-1,1+\frac{\delta}{D}\right)-\zeta\left(\alpha-1,1-\frac{\delta}{D}\right)\right)
,& \mbox{if }\alpha\neq 2.
\end{array} \right.  \label{PFS_avg_interference}
\end{eqnarray}
A simple upper bound is derived by considering the interference
only from the two nearest cells
\begin{eqnarray}
\textsf{C} &\leq& \EE \left[ \log_2\left(1+ \rho s \max_k f_k +\rho
I_0^{(2)} \right)\right]
\nonumber \\
&& - \EE \left[ \log_2 \left( 1+\rho
(g_{\hat{k}(n-1)}(n,n-1)+g_{\hat{k}(n+1)}(n,n+1)) \right) \right].
\label{PFS_upper_bound}
\end{eqnarray}
where the average interference is given by
\begin{equation}
I_0^{(2)}=\frac{D^{-\alpha+1}}{(\alpha-1)(D/2-\delta)} \left(
2^{\alpha-1}-\left(\frac{2}{3}\right)^{\alpha-1}+\left(1+\frac{\delta}{D}\right)^{-\alpha+1}-\left(1-\frac{\delta}{D}\right)^{-\alpha+1}\right).
\end{equation}

The proportional fairness system is also interference limited for
any given finite $K$, because, as SNR increases, the
spectral efficiency converges to a finite value
\begin{equation}
\textsf{C}_0=\lim_{\rho \rightarrow \infty} \textsf{C} = \EE \left[
\log_2\left( \frac{ s  \max_k f_k}{\sum_{j\neq n}
g_{\hat{k}(j)}(n,j)} \right)\right]. \label{PFS_cap_limit}
\end{equation}
However, under mild conditions on $f_k$ (e.g., they are independent and identically distributed (i.i.d.)
exponential random variables), for $K \rightarrow \infty$ we have \cite{Sharif05}
\begin{equation}
\Pr(|\max_k f_k - \log K|< \log\log K) \geq 1-O\left(
\frac{1}{\log K} \right).  \label{mud_gain}
\end{equation}
Consequently, the spectral efficiency limit $\textsf{C}_0$ is
order of $O(\log\log K)$. Notice that the spectral efficiency of the
delay-limited system converges to a finite value $\textsf{C}_0$ in
(\ref{C_0}) as $E_b/N_0\rightarrow \infty$, also in the case of $K \rightarrow \infty$.
We conclude that the PFS delay-tolerant system has merit in terms of spectral efficiency
limit because its spectral efficiency increases without bound as the number of users tends to
infinity (another manifestation of the ubiquitous {\em multiuser diversity} principle).

\begin{figure}
\begin{center}
\includegraphics[width=10cm]{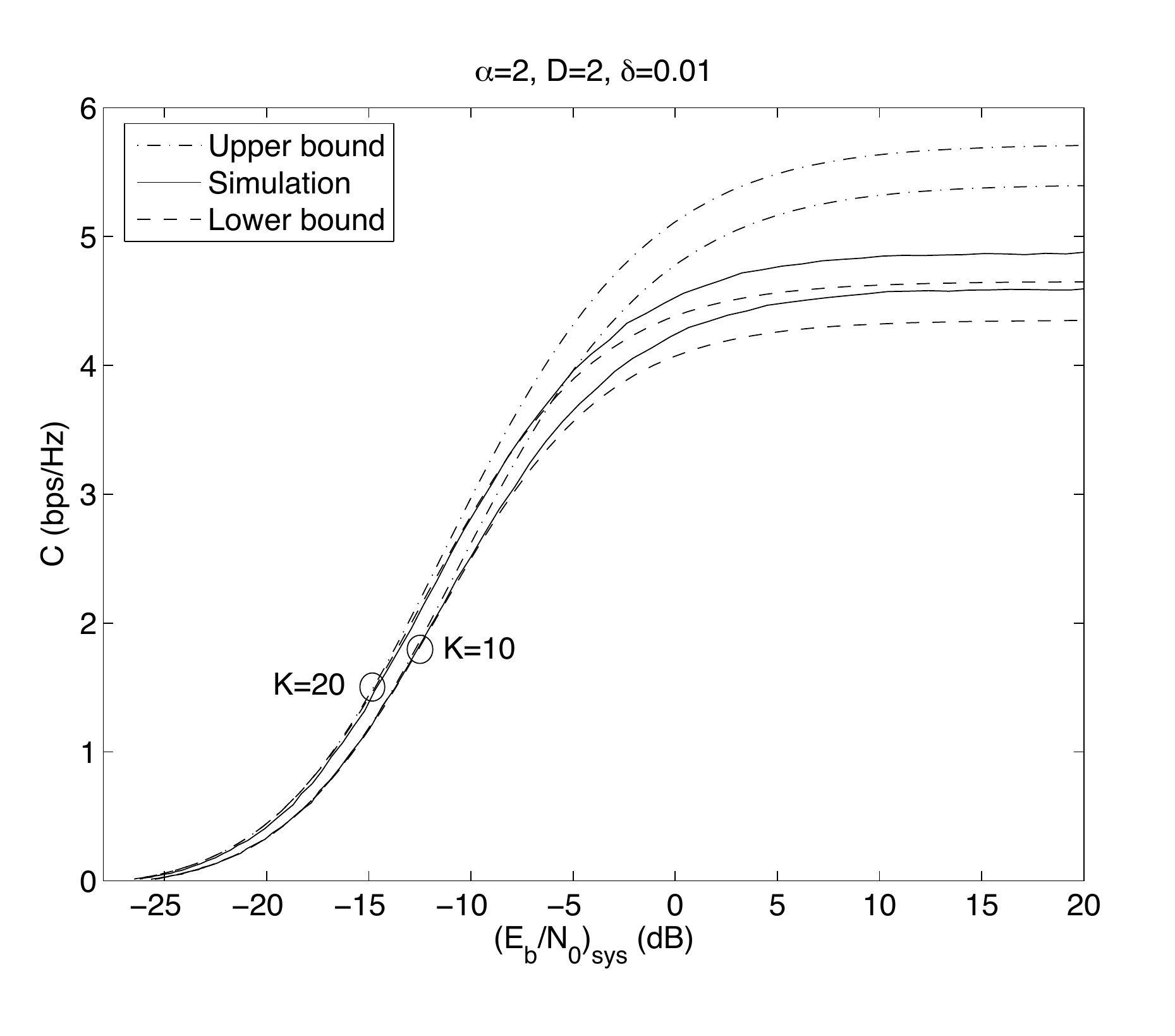}
\end{center}
\caption{Spectral efficiency lower and upper bounds versus system
$E_b/N_0$ for the proportional fair scheduling for $K=10$ and
$K=20$. The channel parameters are path loss exponent $\alpha=2$,
the cell size $D=2$, and the forbidden region $\delta=0.01$.}
\label{Fig_capacity_PFS}
\end{figure}

Fig.~\ref{Fig_capacity_PFS} shows the performance of PFS, the
lower bound (\ref{PFS_lower_bound}), and the upper bound
(\ref{PFS_upper_bound}). The lower bound is quite tight and the
upper bound is close to the simulation result at low SNR regime.
Compared to Fig.~\ref{Fig_capacity_HFS}, we can observe that PFS
generally outperforms the delay-limited system with infinite number of users
in the regime of low to moderate SNR and yields similar throughput at high SNR.

\section{Conclusions}

We provided a closed-form analysis of the system spectral efficiency vs. the system $E_b/N_0$
for two types of systems in a multi-cell ``Wyner-like'' cellular scenario, under the assumption of
conventional single-BS processing. On one hand, we have a hard-fairness system
where users transmit at fixed instantaneous rates in each slot and the system allocates power and makes use of
successive interference cancellation decoding at each BS in order to minimize the required power to
accommodate the user rate requests. On the other hand, we have a delay-tolerant
system with variable instantaneous rate allocation in order to maximize the long-term system throughput
subject to the proportional fairness constraint.
Beyond the pleasant analytical details, the main outcomes of this work are:
1) We validated analytically the popular simplified multi-cell model that treats inter-cell interference power as
proportional to the total cell power, evaluating the proportionality factor $\beta$ and showing that it is indeed close
to a constant almost independent of spectral efficiency;
2) We showed that significant gains can be obtained by optimizing the partial reuse factor, letting users in the outer region of cells
to transmit with duty-cycle 0.5;
3) We showed that PFS yields significant gains at low spectral efficiency, while for a finite number of users and high SNR
the two systems are quite comparable. Hence, there is no silver bullet associated with PFS,
and only a moderate increase in throughput over a hard-fairness system can be expected by exploiting the delay-tolerant nature of data traffic;  4) However, as $K$ increases, PSF yields an unbounded spectral efficiency while the HF
system becomes interference limited.
This is a manifestation of the multiuser diversity in a multi-cell environment.

\begin{center}
\textsc{Appendix A}

\textsc{Proof of Theorem \ref{Theorem1}}
\end{center}

The proof is based on the proof of Theorem 2 in \cite{Caire07}.
Let $R_k^m(n)=\frac{1}{K}\Gamma \nu_k^m (n)$, for all $k$ and $m$,
be the partial rate allocation. We rewrite (\ref{opt_prob}) for
given interference $I^m(n)$ to minimize
\begin{eqnarray}
\frac{1}{N_0 \Gamma} \sum_{k=1}^K \sum_{m=1}^M
\frac{N_0+I^m(n)}{g_{\pi_k^m(n)}^m(n)}
\exp\left(\frac{\Gamma}{K}\sum_{i<k} \nu_{\pi_i^m(n)}^m(n) \right)
\left(\exp\left( \frac{\Gamma}{K} \nu_{\pi_k^m(n)}^m(n) \right) -
1 \right) \label{minimize_energy}
\end{eqnarray}
subject to $\sum_{m=1}^M \nu_k^m(n) = \nu_k(n)$, $k=1,2,\cdots,K$,
and to the non-negative constraints $\nu_k^m(n)\geq 0$ for all $k$
and $m$. Since $\exp(x)=1+x+o(x)$ for small $x$, the objective
function for large $K$ can be written as
\begin{eqnarray}
\frac{1}{N_0} \frac{1}{K} \sum_{k=1}^K \sum_{m=1}^M (N_0+I^m(n))
\frac{\nu_{\pi_k^m(n)}^m(n)}{g_{\pi_k^m(n)}^m(n)}
\exp\left(\frac{\Gamma}{K}\sum_{i<k} \nu_{\pi_i^m(n)}^m(n) \right)
\label{minimize_energy2}
\end{eqnarray}
and its associated Lagrangian function is
\begin{eqnarray}
\mathcal{L}=\frac{1}{N_0} \frac{1}{K} \sum_{k=1}^K \sum_{m=1}^M
\frac{\nu_{\pi_k^m(n)}^m(n)}{d_{\pi_k^m(n)}^m(n)}
\exp\left(\frac{\Gamma}{K}\sum_{i<k} \nu_{\pi_i^m(n)}^m(n) \right)
- \sum_{k=1}^K \lambda_k^{(K)}\left( \sum_{m=1}^M \nu_k^m-\nu_k
\right),
\end{eqnarray}
where the channel SINR (signal to interference plus noise ratio)
$d_{\pi_k^m(n)}^m(n)$ is given by $d_{\pi_k^m(n)}^m(n)\equiv
g_{\pi_k^m(n)}^m(n)/(N_0+I^m(n))$. We have an optimization problem
for each $K$ and $\lambda_k^{(K)}$ denotes the $k$th Lagrangian
multiplier of the problem. By differentiating with respect to
$\nu_k^m$, and by letting $\pi_i^m=k$, i.e., user $k$ is ranked in
the $i$th position on subchannel $m$, the Kuhn-Tucker condition
becomes
\begin{eqnarray}
\frac{1}{d_{k}^m(n)}\exp\left(\frac{\Gamma}{K}\sum_{l<i}\nu_{\pi_l^m(n)}^m(n)\right)
+\frac{\Gamma}{K}\sum_{l>i}
\frac{\nu_{\pi_l^m(n)}^m(n)}{d_{\pi_l^m(n)}^m(n)}
\exp\left(\frac{\Gamma}{K}\sum_{l'<l} \nu_{\pi_{l'}^m(n)}^m(n)
\right) \geq \lambda_k^{(K)}, \label{KKT_condition}
\end{eqnarray}
where we have multiplied both sides by $KN_0$ and have replaced
$KN_0\lambda_k^{(K)}$ by $\lambda_k^{(K)}$. At this point, we notice that
(\ref{KKT_condition}) coincides with
the Kuhn-Tucker condition for the single cell case (see proof of Theorem 2 in \cite{Caire07})
provided that SNR \cite{Caire07} is replaced by
the SINR $g_{\pi_k^m(n)}^m(n)/(N_0+I^m(n))$.
Consequently, the asymptotically optimal rate allocation for large
$K$ is given by \cite{Caire07}
\begin{equation}
\nu_k^m(n) = \left\{ \begin{array}{cc}
  \nu_k(n), & \mbox{for}~m=m_k \\
  0, & \mbox{for}~m\neq m_k
\end{array} \right.  \label{optimal_nu}
\end{equation}
where $m_k=\arg\max_l \{g_k^l(n,n)/(N_0+I^m(n))\}$ denotes the
index of the subchannel for which user $k$ has {\em maximum SINR}.
As a byproduct, we have that allocating each user to its own
best  SINR channel is asymptotically optimal for large $K$ in the multiple cell case.
Since the subchannel distribution is symmetric and
users are distributed uniformly in each cell, the interference
$I^m(n)$ converges to the same non-random limit for all $m$ and $n$ as $K \rightarrow \infty$.
If the interference is the same, then the optimal rate allocation for the multiple
cell consists of allocating each user on its best SNR channel, i.e.,
$m_k=\arg\max_l \{g_k^l(n,n)\}$.

Now we show that allocating each user on its subchannel with best channel
gain  makes indeed the interference $I^m(n)$ constant with $m$ (constant with the cell index $n$ follows immediately form the symmetry of the system). From (\ref{interference}), for large $K$, the
interference power is expressed as
\begin{eqnarray}
I^m(n) = \sum_{j\neq n} (N_0+I^m(j)) \frac{\Gamma}{K} \sum_{k=1}^K
\frac{g_{\pi_k^m(n)}^m(n,j)
\nu_{\pi_k^m(j)}^m(j)}{g_{\pi_k^m(j)}^m(j,j)}
 \exp\left(\frac{\Gamma}{K}\sum_{i<k}
\nu_{\pi_i^m(j)}^m(j) \right) + o(1/K). \label{inteference_exp}
\end{eqnarray}
As we adopt an optimal successive decoding scheme in each BS,
$\nu_k^m(j)$ takes the form in (\ref{optimal_nu}). Suppose that $g_k^m(j,j)$,
given by $s_k(j,j)=x$ and $f_k^m(j,j)=y$, is ranked in the $i$th
position by the permutation $\pi^m(j)$. By using
Lemma~\ref{lemma1} in Appendix B, we get
\begin{equation}
\exp\left(\frac{\Gamma}{K}\sum_{i<k} \nu_{\pi_i^m(j)}^m(j) \right)
\rightarrow \exp\left(\frac{\Gamma}{M} G_M(xy)
\right)
\end{equation}
as $K\rightarrow \infty$.
From (\ref{s_k(n,j)}), averaging with respect to $\theta_k(n,j)$, we get
\begin{equation}
\EE \left[ s_k(n,j) | s_k(j,j) \right] = \frac{1}{2} \left(
\left(|n-j|D-s_k(j,j)^{-\frac{1}{\alpha}}\right)^{-\alpha} +
\left(|n-j|D+s_k(j,j)^{-\frac{1}{\alpha}}\right)^{-\alpha} \right)
\end{equation}
According to Lemma~\ref{lemma2} in Appendix B, the asymptotic
interference in (\ref{inteference_exp}) for large $K$ becomes
\begin{eqnarray}
I^m(n) = \sum_{j\neq n} (N_0+I^m(j)) \frac{\Gamma}{M} \iint
e^{\frac{\Gamma}{M} G_M(xy)} \Phi(x,|n-j|D)
\frac{dF_s(x)dH_M(y)}{xy},~ \label{interference_relation}
\end{eqnarray}
where $\Phi(x,y)\equiv \frac{1}{2} \left(
\left(y-x^{-\frac{1}{\alpha}}\right)^{-\alpha} +
\left(y+x^{-\frac{1}{\alpha}}\right)^{-\alpha}\right)$. Note that
the interference relation in (\ref{interference_relation}) is
symmetric with respect to $I^m(n)$ for all $m$ and $n$. Therefore,
as $K\rightarrow \infty$, the interference at each base station
converges to the same value, $I^m(n)\rightarrow I_0$, for all $m$
and $n$. As $I^m(n)\rightarrow I_0$, we have
\begin{eqnarray}
I_0 = (N_0+I_0) \frac{\Gamma}{M} \iint e^{\frac{\Gamma}{M}
G_M(xy) } \Phi(x)
\frac{dF_s(x)dH_M(y)}{xy},
\label{interference_relation2}
\end{eqnarray}
where $\Phi(x) \equiv 2 \sum_{j=1}^{\infty} \Phi(x,jD)$. Solving
(\ref{interference_relation2}) with respect to $I_0$, we obtain
the asymptotic interference under the single-cell optimal
successive decoding strategy. In addition, by Lemma~\ref{lemma1}
in Appendix B, allocating each user on its own best channel makes
the minimum $(E_b/N_0)_{\rm sys}$ in (\ref{minimize_energy2})
converges to the following
\begin{equation}
\left( \frac{E_b}{N_0} \right)_{\rm sys} \rightarrow
\frac{N_0+I_0}{N_0} \int_0^{\infty} \frac{1}{x} \exp\left(
\frac{\Gamma}{M} G_M(x) \right)dG_M(x)
\label{EbN0_int}
\end{equation}
as $K\rightarrow \infty$ (again, this can be easily shown based on
the proof of Theorem 2 in \cite{Caire07}). Inserting the
expression of $I_0$ from (\ref{interference_relation2}) into
(\ref{EbN0_int}) and expressing $\Gamma$ in bits, we eventually
arrive at the desired result.

\begin{center}
\textsc{Appendix B}

\textsc{Lemmas}
\end{center}

\begin{lemma} \cite{Caire07} \label{lemma1}
Let $\textsl{A}$ be an interval, and $\textsl{A}(m)$ denote the
set
\begin{equation}
\textsl{A}(m)=\left
\{k|s_k(n,n)\max\{f_k^1(n,n),\cdots,f_k^M(n,n)\}\in \textsl{A},
m_k=m \right \},
\end{equation}
where $m_k\equiv \arg\max_l \{f_k^l(n,n)\}$. Also, let $g(x)$
denote a continuous measurable function in $x\in A$. Under
Assumptions A2, A3, and A4, we get
\begin{equation}
\frac{1}{K}\sum_{k\in \textsl{A}(m)} g(s_k(n,n) f_k^m(n,n))  \nu_k
\rightarrow \frac{1}{M}\int_{x\in\textsl{A}} g(x)
dG_M(x)
\end{equation}
with probability 1, as $K\rightarrow \infty$ and $M$ is fixed.
\end{lemma}


\begin{lemma} \label{lemma2}
Let $\textsl{A}$ be an interval, and $\textsl{A}(m)$ denote the
set
\begin{equation}
\textsl{A}(m)=\left
\{k|s_k(j,j)\max\{f_k^1(j,j),\cdots,f_k^M(j,j)\}\in \textsl{A},
m_k=m \right \},
\end{equation}
where $m_k\equiv \arg\max_l \{f_k^l(j,j)\}$. Also, let
$g(x,\phi)$ denote a continuous measurable function in $x\in A$
and $\phi \geq 0$. Under Assumptions A2, A3, and A4, we get
\begin{equation} \label{bubu}
\frac{1}{K}\sum_{k\in \textsl{A}(m)}
\frac{g(s_k(j,j),\theta_k(n,j)) f_k^m(n,j) \nu_k}{s_k(j,j)
f_k^m(j,j)} \rightarrow \frac{1}{M}\iint_{xy\in\textsl{A}}
\frac{\EE_{\theta}[g(x,\theta)]}{xy} dF_s(x)dH_M(y)
\end{equation}
with probability 1, as $K\rightarrow \infty$ and $M$ is fixed.
In (\ref{bubu}), $\EE_\theta[\cdot]$ denotes expectation with respect to
$\theta \sim F_\theta(\phi)$.
\end{lemma}
\begin{proof}
It follows from the convergence of the empirical cdfs that, as
$K\rightarrow \infty$, we have
\begin{eqnarray}
\frac{1}{K}\sum_{k\in \textsl{A}(m)}
\frac{g(s_k(j,j),\theta_k(n,j)) f_k^m(n,j) \nu_k}{s_k(j,j)
f_k^m(j,j)}~~~~~~~~~~~~~~~~~~~~~~~~~~ \nonumber \\ =
\frac{|\textsl{A}(m)|}{K} \frac{1}{|\textsl{A}(m)|} \sum_{k\in
\textsl{A}(m)} \frac{g(s_k(j,j),\theta_k(n,j))
f_k^m(n,j) \nu_k}{s_k(j,j) f_k^m(j,j)}~~~~~~~~~~ \nonumber \\
\rightarrow \PP(k\in \textsl{A}(m))
\EE\left[\frac{g(s_k(j,j),\theta_k(n,j)) f_k^m(n,j) \nu_k}{s_k(j,j)
f_k^m(j,j)} | k\in \textsl{A}(m) \right].
\end{eqnarray}
We have
\begin{eqnarray}
\PP(k\in \textsl{A}(m)) &=&
\PP(s_k(j,j)\max\{f_k^1(j,j),\cdots,f_k^M(j,j)\}\in \textsl{A}|
m_k=m)\PP(m_k=m) \nonumber \\ &=&\frac{1}{M}\int_{\textsl{A}}
dG_M(x). \label{prob1}
\end{eqnarray}
where $G_M(x)$ is the limit empirical cdf of $s_k(j,j)\max\{f_k^1(j,j),\cdots,f_k^M(j,j)\}$, which exists a.s..
By Assumptions A3 and A4, we also have
\begin{eqnarray}
\EE\left[\left . \frac{g(s_k(j,j),\theta_k(n,j)) f_k^m(n,j) \nu_k}{s_k(j,j)
f_k^m(j,j)} \right | k\in \textsl{A}(m) \right]~~~~~~~~~~~~~~~~~~~~~~~~~~~~~~~~~~~~~~~~~~~~~~~~~ \nonumber \\
= \EE\left[\left . \frac{g(s_k(j,j),\theta_k(n,j))}{s_k(j,j) f_k^m(j,j)} \right |
k\in \textsl{A}(m) \right] \EE\left[ f_k^m(n,j) \nu_k | k\in
\textsl{A}(m) \right]~~~~~~~~~~~~~~~~~~~~~~~~~~ \nonumber \\
= \EE\left[\frac{g(s_k(j,j),\theta_k(n,j))}{s_k(j,j)
\max\{f_k^1(j,j),\cdots,f_k^M(j,j)\}} | s_k(j,j)
\max\{f_k^1(j,j),\cdots,f_k^M(j,j)\} \in \textsl{A} \right]
\nonumber \\
=\frac{1}{\int_{\textsl{A}} dG_M(x)} \iint_{xy\in
\textsl{A}} \frac{\EE_{\theta}[g(x,\theta)]}{xy}
dF_s(x)dH_M(y)~~~~~~~~~~~~~~~~~~~~~~~~~~~~
\label{prob2}
\end{eqnarray}
where $H_M(y)$ denotes the limit empirical cdf of $\max\{f_k^1(j,j),\cdots,f_k^M(j,j)\}$.
By using (\ref{prob1}) and (\ref{prob2}) the result follows.
\end{proof}

\begin{center}
\textsc{Appendix C}

\textsc{Proof of Theorem \ref{Theorem2}}
\end{center}

We can write the minimization of $(E_b/N_0)_{\rm sys}$ for given
interference $I^m$ as minimize
\begin{eqnarray}
\frac{N_0+I^m}{N_0 \Gamma} \sum_{k=1}^K \sum_{m=1}^M
\frac{1}{g_{\pi_k^m}^m} \exp\left(\frac{\Gamma}{K}\sum_{i<k}
\nu_{\pi_i^m}^m \right) \left(\exp\left( \frac{\Gamma}{K}
\nu_{\pi_k^m}^m \right) - 1 \right) \label{simple_minimize_energy}
\end{eqnarray}
subject to $\sum_{m=1}^M \nu_k^m = \nu_k$, $k=1,2,\cdots,K$, and
to the non-negative constraints $\nu_k^m\geq 0$ for all $k$. As
done in Appendix A, the solution to the problem
(\ref{simple_minimize_energy}) for given interference level $I^m$
is the same with the solution of Theorem 2 in \cite{Caire07}.
Consequently, the asymptotically optimal rate allocation for large
$K$ is given by (\ref{optimal_nu}). In addition, by
Lemma~\ref{lemma1} in Appendix B, for given $I^m$, the minimum
$(E_b/N_0)_{\rm sys}$ converges to the following
\begin{equation}
\left( \frac{E_b}{N_0} \right)_{\rm sys} \rightarrow
\frac{N_0+I^m}{N_0} \int \frac{1}{x} \exp\left( \frac{\Gamma}{M}
G_M(x) \right)dG_M(x)
\label{Simple_EbN0_int}
\end{equation}
Inserting (\ref{simple_opt_energy}) into
(\ref{simple_interference}) and applying Lemma~\ref{lemma1} in
Appendix B once again, we get
\begin{eqnarray}
I^m = \beta(N_0+I^m) \int \frac{1}{x} \exp\left(\frac{\Gamma}{M}
G_M(x)\right) dG_M(x)
\end{eqnarray}
as $K\rightarrow \infty$. Since this equation holds for all $m$,
$I^m$ converges to the same value $I_0$ independent of $m$.
Consequently, the minimum $(E_b/N_0)_{\rm sys}$ converges to the
following
\begin{equation}
\left( \frac{E_b}{N_0} \right)_{\rm sys} \rightarrow \frac{ \int
\frac{1}{x} e^{\frac{\Gamma}{M} G_M(x)}
dG_M(x)} {1- \beta \frac{\Gamma}{M} \int \frac{1}{x}
e^{\frac{\Gamma}{M} G_M(x)} dG_M(x)}
\label{Simple_EbN0_int2}
\end{equation}
as $K\rightarrow \infty$. Expressing $\Gamma$ in bits yields the
desired result.

\begin{center}
\textsc{Appendix D}

\textsc{Proof of Theorem \ref{Theorem_HF_Partial}}
\end{center}

The proof is very similar to the proof of Theorem \ref{Theorem1}.
Therefore, we give only a sketch and leave out trivial details for
the sake of space limitation. From (\ref{C_partial}), the numbers
of bits transmitted in each even cell and each odd cell are
$\Gamma_0=\frac{2M\textsf{C}(1-\delta)}{1+r_0-2\delta}$ and
$\Gamma_1=\frac{2M\textsf{C}(r_0-\delta)}{1+r_0-2\delta}$. From
(\ref{inteference_exp}), the interference experienced by cell $n$
is given by
\begin{eqnarray}
I^m(n) = \sum_{j\neq n,~{\rm even}~j} (N_0+I^m(j))
\frac{\Gamma_0}{K_0} \sum_{k=1}^{K_0} \frac{g_{\pi_k^m(n)}^m(n,j)
\nu_{\pi_k^m(j)}^m(j)}{g_{\pi_k^m(j)}^m(j,j)}
 \exp\left(\frac{\Gamma_0}{K_0}\sum_{i<k}
\nu_{\pi_i^m(j)}^m(j) \right)~~~~~~~~  \nonumber \\
+ \sum_{j\neq n,~{\rm odd}~j} (N_0+I^m(j)) \frac{\Gamma_1}{K_1}
\sum_{k=1}^{K_1} \frac{g_{\pi_k^m(n)}^m(n,j)
\nu_{\pi_k^m(j)}^m(j)}{g_{\pi_k^m(j)}^m(j,j)}
 \exp\left(\frac{\Gamma_1}{K_1}\sum_{i<k}
\nu_{\pi_i^m(j)}^m(j) \right)  + o(1/K)
\end{eqnarray}
The asymptotically optimal rate allocation is to allocate each
user on its best SNR channel as done in Theorem~\ref{Theorem1}.
According to Lemma~\ref{lemma2} in Appendix B, as $K\rightarrow
\infty$, the asymptotic interference converges to
\begin{eqnarray}
& I^m(n) = \sum_{j\neq n,~{\rm even}~j} (N_0+I^m(j))
\frac{\Gamma_0}{M} \iint e^{\frac{\Gamma_0}{M}
G_M(xy)} \Phi(x,2|n-j|)
\frac{dF_s(x)dH_M(y)}{xy} & \nonumber \\
& +
\sum_{j\neq n,~{\rm odd}~j} (N_0+I^m(j)) \frac{\Gamma_1}{M} \iint
e^{\frac{\Gamma_1}{M} G_M^{(\delta,r_0)}(xy)}
\Phi(x,2|n-j|)
\frac{dF_s^{(\delta,r_0)}(x) dH_M(y)}{xy}, &
\label{int_partial}
\end{eqnarray}
where $\Phi(x,y)\equiv \frac{1}{2} \left(
\left(y-x^{-\frac{1}{\alpha}}\right)^{-\alpha} +
\left(y+x^{-\frac{1}{\alpha}}\right)^{-\alpha}\right)$. Due to its
symmetry, the interference power at each even cell converges to
the same limit $I_0$, and the interference power to all odd cells
converges to the same limit $I_1$. Now we define $\Phi_0(x)$ and
$\Phi_1(x)$ to be $\Phi_0(x)\equiv 2\sum_{j=2,4,\cdots}
\Phi(x,2j)$ and $\Phi_1(x)\equiv 2\sum_{j=1,3,\cdots} \Phi(x,2j)$.
Then, we can rearrange (\ref{int_partial}) to get
\begin{eqnarray}
I_0&=&(N_0+I_0)A_{00}+(N_0+I_1)A_{01} \nonumber \\
I_1&=&(N_0+I_0)A_{10}+(N_0+I_1)A_{11}, \label{I_0,I_1_eq}
\end{eqnarray}
where $A_{00},A_{01},A_{10},A_{11}$ are given by (\ref{eq30a}) --
(\ref{eq30d}), respectively. Solving for $I_0,I_1$ in terms of the
$A_{i,j}$'s and expressing the limiting total energy in each even
and odd cell using Lemma 1 in Appendix B, we finally get the
desired result.

\begin{center}
\textsc{Appendix E}

\textsc{Derivation of lower and upper bounds for PFS}
\end{center}

We derive the lower bound using Jensen's inequality. The
average interference can be obtained by
\begin{eqnarray}
I_0 = 2 \sum_{j=1}^{\infty}
\EE\left[\frac{1}{2}(jD-U)^{-\alpha}+\frac{1}{2}(jD+U)^{-\alpha}
\right]~~~~~~~~~~~~~~~~~~~~~~~~~~~~~~~~~~~~~~~
\nonumber \\
= \frac{D^{-\alpha+1}}{(\alpha-1)(D/2-\delta)} \left(2^{\alpha-1}+
\sum_{j=1}^{\infty} \left( \frac{1}{(j+\delta/D)^{\alpha-1}}
-\frac{1}{(j-\delta/D)^{\alpha-1}} \right) \right),~~~~~
\end{eqnarray}
where we used $D=2r$. From the identity \cite{Gradshteyn00}
\begin{equation}
\sum_{j=1}^{\infty} \frac{1}{j^2-x^2}=\frac{1}{2x^2}(1-\pi x
\cot(\pi x))
\end{equation}
for $\alpha=2$, then the average interference is given by
(\ref{PFS_avg_interference}). Because $\log(1+x/(1+y))$ is convex
with respect to $y$, Jensen's inequality yields the lower bound in
(\ref{PFS_lower_bound}).

For the upper bound, we only consider the interference from two
nearest cells
\begin{eqnarray}
C &\leq& \EE \left[ \log_2\left(1+ \frac{\rho s  \max_k f_k}{1+\rho
(g_{\hat{k}(n-1)}(n,n-1)+g_{\hat{k}(n+1)}(n,n+1)) }
\right)\right] \nonumber \\
 &\leq& \EE \left[ \log_2\left(1+ \rho s
\max_k f_k +\rho I_0^{(2)} \right)\right] \nonumber \\
&& - \EE \left[ \log_2 \left( 1+\rho
(g_{\hat{k}(n-1)}(n,n-1)+g_{\hat{k}(n+1)}(n,n+1)) \right) \right],
\end{eqnarray}
where the last inequality follows from Jensen's inequality and the
average interference from the two nearest cells is given by
\begin{eqnarray}
I_0^{(2)} = \EE[g_{\hat{k}(n-1)}(n,n-1)+g_{\hat{k}(n+1)}(n,n+1))]~~~~~~~~~~~~~~~~~~~~~~~~~~~~~~~~~~~~~~~~~~
\nonumber \\
= \frac{D^{-\alpha+1}}{(\alpha-1)(D/2-\delta)} \left(
2^{\alpha-1}-\left(\frac{2}{3}\right)^{\alpha-1}+\left(1+\frac{\delta}{D}\right)^{-\alpha+1}-\left(1-\frac{\delta}{D}\right)^{-\alpha+1}\right).
\end{eqnarray}

\newpage


\newpage

\end{document}